\documentclass[superscriptaddress,pra,twocolumn,showpacs,amsmath,amssymb,aps,a4paper]{revtex4}

 \usepackage{graphicx}
 \usepackage{dcolumn}
 \usepackage{bm}
 \usepackage{amssymb,amsmath}
 \usepackage{natbib}
 \usepackage{verbatim}
 \usepackage{subfigure}
 \usepackage{textcomp}

\newcommand{\PRYSO}{Pr$^{3+}$:Y$_2$SiO$_5$ }
\usepackage{cancel}

\begin{document}

  \title{Photon echo quantum memories in inhomogeneously broadened two level atoms}

 \author{D. L. McAuslan}
 \affiliation{Jack Dodd Centre for Photonics and Ultra-Cold Atoms, Department of Physics,
University of Otago, Dunedin, New Zealand.}
 \author{P. M. Ledingham}
 \affiliation{Jack Dodd Centre for Photonics and Ultra-Cold Atoms, Department of Physics,
University of Otago, Dunedin, New Zealand.}
 \author{W. R. Naylor}
 \affiliation{Jack Dodd Centre for Photonics and Ultra-Cold Atoms, Department of Physics,
University of Otago, Dunedin, New Zealand.}
 \author{S. E. Beavan}
 \affiliation{Centre for Quantum Computation and Communication Technology, RSPE, Australian National University, Canberra, ACT 0200, Australia} 
 \author{M. P. Hedges}
 \affiliation{Centre for Quantum Computation and Communication Technology, RSPE, Australian National University, Canberra, ACT 0200, Australia} 
 \author{M. J. Sellars}
 \affiliation{Centre for Quantum Computation and Communication Technology, RSPE, Australian National University, Canberra, ACT 0200, Australia}  
\author{J. J. Longdell}
 \email{jevon.longdell@otago.ac.nz}
 \affiliation{Jack Dodd Centre for Photonics and Ultra-Cold Atoms, Department of Physics,
University of Otago, Dunedin, New Zealand.}

\date{\today}
\begin{abstract}
  Here we propose a solid-state quantum memory that does not require spectral
  holeburning, instead using strong rephasing pulses like traditional
  photon echo techniques.  The memory uses external
  broadening fields to reduce the optical depth and so switch off the
  collective atom-light interaction when desired. The proposed memory
  should allow operation with reasonable efficiency in a much broader
  range of material systems, for instance Er$^{3+}$ doped crystals which have a transition at 1.5~$\mu m$. We present analytic theory supported by
  numerical calculations and initial experiments.
\end{abstract}

\pacs{3.67.Lx,82.53.Kp,78.90.+t}
   \keywords{quantum memory, photon echo, rare earth}
\maketitle

\section{Introduction}

Photon echo based techniques, particularly in rare earth ion dopants,
have been long investigated for classical signal processing
\cite{moss82}.  More recently they have found use in quantum
memory applications and are now leading the field with the
longest storage times \cite{frav05,stopped}, the highest efficiencies
\cite{hedg10}, broadest bandwidths \cite{sagl10} and the highest
time-bandwidth products \cite{usma10}.  The photon echo techniques
used for quantum information applications differ from those
used in the classical domain.  Classical information processing
using rare earth ion dopants relies on strong optical pulses to cause
rephasing of the atomic coherences. Applying a similar approach in the quantum realm would be
highly desirable, but analysis has shown that the
basic techniques, namely the two-pulse \cite{ledi10,rugg08} and three-pulse \cite{sang10} photon echo sequences, will not work as quantum memories.

This work proposes low noise photon echo techniques based on optical
rephasing pulses that could be useful as quantum memories. We
present analytical theory and initial experiments.

The two approaches currently used for photon echo quantum memories
techniques, controlled reversible inhomogeneous broadening
(CRIB)\cite{mois01,krau06,alex06,long08,hete08,hedg10} and atomic
frequency combs (AFC)\cite{ried08,afze09} both require spectral
holeburning. This is the process where the inhomogeneously broadened
absorption profile is modified by frequency selective optical pumping
to shelving states.

For CRIB, spectral holeburning is used to create a spectrally narrow
absorption feature.  This is then broadened with a field gradient to
accept the input light. The inverse spectral-width of the initial
feature determines the storage time, and how far it is broadened
determines the
bandwidth. The light is then recalled by reversing the field
gradient. To do this with high efficiency, significant optical
thickness is required for the broadened feature, which means that
extremely large optical thicknesses are required for the unbroadened
feature. Hedges et al.\cite{hedg10} started with an absorption of
-130~dB in the initial feature to achieve 69\% efficiency with a
time-bandwidth product only large enough to faithfully store one
pulse.

In AFC protocols the material is prepared in a similar way to CRIB except that
a number of narrow absorption features are created. Using this technique
large bandwidth delays \cite{usma10} and reasonable efficiencies have also been
demonstrated \cite{chan10}.

With both CRIB and AFC methods, high efficiencies require efficient
optical pumping and this requires long-lived shelving states. To date
this has meant using shelving states in the dopant ions' hyperfine
structure.  Our proposal does not rely on spectral holeburning, and
the removal of this restriction is very appealing. The
hyperfine splittings are generally less than 100~MHz, which makes
large bandwidth operation for the AFC and CRIB protocols
problematic. Saglamyurek et al. \cite{sagl10} used a trick to overcome
the problem of hyperfine structure and bandwidth with AFC memories. By
choosing the correct comb spacing, they had the anti-holes that result 
from spectral holeburning appearing on top of the comb teeth. The downside
of this approach is that high finesse frequency combs are not possible,
severely limiting the efficiency.  

Another reason to avoid
spectral holeburning is the requirement of efficient holeburning
limits the choice of material systems. For example, while single
photon memories have been demonstrated at telecommunication
wavelengths using erbium \cite{laur10}, the lack of an efficient
holeburning mechanism severely limited the memory efficiency.


Ensemble quantum memories rely on the fact that the collective
interaction with an optical field is greatly enhanced for systems
with reasonable optical depths. The problem with optically rephased
memories is that the strong optical rephasing pulses necessarily invert
the population, this causes gain and the associated noise. 

Here we propose avoiding this problem by using a perturbing field to
spectrally broaden the ensemble while in the excited state to the
point where it is optically thin. This removes the collective
enhancement and stops the unwanted noise processes. We call this
method where both reversible inhomogeneous broadening and optical
rephasing methods are used hybrid photon echo rephasing (HYPER).

Another approach that avoids the use of spectral holeburning by using
Raman echoes has also been proposed \cite{mois11}.

\section{Description}

The hybrid photon echo rephasing technique uses a combination of
broadening field pulses and optical pulses to rephase a small optical
input pulse. For what follows we shall assume that the broadening
field is an electric field gradient which causes inhomogeneous
broadening due to the atom's linear Stark effect. Using a linear
electric field gradient is not a fundamental requirement of the HYPER
protocol. As long as the broadening is completely reversed any applied
field will be acceptable. The case of a linear field gradient is
considered here as this is what was used in the experimental work.

Using two $\pi$-pulses allows the ions to rephase near the ground
state, forming an echo of the normal two pulse echo (2PE). This is
beneficial as an echo forming in the excited state is inherently
noisy. The electric field gradients are applied at appropriate times
to eliminate excited state noise on the output.

\begin{figure}
  \centering
  \includegraphics[width=0.48\textwidth]{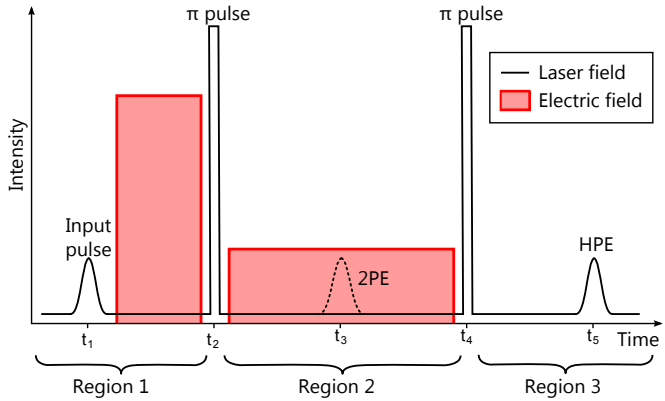}
\caption{\label{fig:pulseseq} (Color online)
Sequence of pulses used to create a hybrid echo. The dashed line shows
where the 2PE would form if there was no electric field applied to the
sample. Also shown are the three time regions used when we explain
the echo sequence.
}
\end{figure}

The pulse sequence, shown in Fig.~\ref{fig:pulseseq}, has three time
regions. In the first region ($0 < t < t_2$) one or more input pulses are
applied followed by the application of a broadening field. A $\pi$
pulse is applied at $t_2$.  The $\pi$-pulse creates  gain 
which a 2PE would ordinarily propagate through and thus undergo 
amplification. The effect of broadening the ensemble is to reduce the gain, and
the two pulse echo, to an arbitrarily low level.  In doing so the 
troublesome collective interaction between the light and the atoms is
effectively turned off while the atoms are in the excited state.

While the dephasing caused by the applied broadening stops the two
pulse echo forming, so long as the broadening applied before and after the
first $\pi$-pulse are balanced an echo of the 2PE will form after a
second $\pi$-pulse. This is the output from the memory. We call this
type of echo, which involves both static and controlled inhomogeneous
broadening a hybrid photon echo (HPE).

\section{Analytical Theory}

Here we use the same theoretical description that was developed for rephased
amplified spontaneous emission \cite{ledi10}. It is assumed that the input
light to all regions is weak compared to a $\pi$-pulse and that the
rephasing pulses are ideal $\pi$-pulses.

Figure \ref{fig:pulseseq} shows the pulse sequence with three different
time regions. The equations of motion for time region 1 before the
Stark shift is applied are as follows,

\begin{subequations}
 \label{equ:groundMBE}
 \begin{align}
  \label{equ:MBE1R1}
  \frac{\partial}{\partial t} & \hat{D}_{1 }(z,t,\Delta) = i \Delta \hat{D}_{1} \; +  \;\; i \hat{a}_{1} \\
  \label{equ:MBE2R1}
  \frac{\partial}{\partial z} & \hat{a}_{1}(z,t) = \frac{i \alpha}{2 \pi} \int_{-\infty}^{\infty} d\Delta \;   \hat{D}_{1} \;.
 \end{align}
 \end{subequations}

 Here, $\hat{a} (\hat{D})$ is the quantum optical (atomic) field
 operator, subscript 1 indicates region 1, $\alpha$ is the optical
 depth parameter and $\Delta$ represents the detuning of the
 atom. The solution for the
 atomic field (Eq.~\ref{equ:MBE1R1}) is given by
\begin{equation}
\hat{D}_1(z,t,\Delta) = i \int^{t}_{-\infty} \hat{a}_1(z,t') \text{e}^{i\Delta (t-t')} dt' + \hat{D}(z,t_0,\Delta)\text{e}^{i\Delta t }\label{equ:D1solution}
\end{equation}
To obtain the solution for the optical field we Fourier transform
Eq.~\ref{equ:D1solution} with respect to time and substitute into the
transformed version of Eq.~\ref{equ:MBE2R1}. The solution in the time
domain is found to be
\begin{equation}
\hat{a}_1(z,t) = \hat{a}_1(0,t)\text{e}^{-\frac{\alpha z}{2}} +
i\alpha \int^z_0 dz' \hat{\mathcal{D}}_1(z',t)\text{e}^{-\frac{\alpha}{2}(z-z')}\label{equ:a1solution}
\end{equation}
where $\hat{\mathcal{D}}_1(z,t)$ is defined as
$\mathcal{F}^{-1}\left[\hat{D}_1(z,t = 0,\Delta=\omega)\right]$. We now have
the complete solution in region 1 for before the Stark shifting fields
are applied.

We now assume that the Stark shifting field is turned on just before
the first $\pi$-pulse and is strong and temporally short, allowing the
dynamics of the optical field to be ignored. The Stark shifting field
alters an atom's detuning dependent on its position by $\eta z$ such
that $\Delta \to \Delta + \eta z$, whereas the $\pi$-pulse inverts the
atoms leading to $\hat{D} \to \hat{D}^\dagger$. The Stark
shift and $\pi$-pulse result in the following transformation on the
atomic field at the region 2 ($t = t_2$) boundary,
\begin{equation}
\hat{D}^{\dagger}_2(z,t = t_2,\Delta) = \left[ \hat{D}_1(z,t_2,\Delta) \text{e}^{i\eta_1z} \right]^{\dagger}
\end{equation}
where $\eta_1 = \eta (t_f -t_i)$. Here, $\eta$ is the intensity of the Stark shifting field and $t_f - t_i$ is the duration of that field.

Under the small input pulse approximation the atoms in region 2 are
all near the excited state.

 A temporally long Stark shifting pulse is applied during all of
 region 2. The equations of motion are given by
\begin{subequations}
 \label{equ:excitedMBE}
 \begin{align}
  \label{equ:MBE1R2}
  \frac{\partial}{\partial t} & \hat{D}_{2}^{\dagger}(z,t,\Delta) = i (\Delta + \eta' z) \hat{D}_{2}^{\dagger} \; -  \;\; i \hat{a}_{2} \\
  \label{equ:MBE2R2}
  \frac{\partial}{\partial z} & \hat{a}_{2}(z,t) = \frac{i \alpha}{2 \pi} \int_{-\infty}^{\infty} d\Delta \;   \hat{D}_{2}^{\dagger} \;,
 \end{align}
 \end{subequations}
 where again the subscript denotes region. It is noted that these equations are similar in form to the region 1 equations, with the atomic fields daggered due to the $\pi$-pulse and the detuning is shifted with a field intensity $\eta'$ such that $\Delta \to \Delta + \eta' z$. Also, a sign change in front of the optical field in the atomic equation of motion is present.
 
 The atomic field solution for region 2 is given by
 \begin{align}
 \hat{D}^\dagger_2(z,t,\Delta) &= -i\int^t_{-\infty}\hat{a}_2(z,t')\text{e}^{i(\Delta+\eta' z)(t-t')}\notag \\
 & \qquad+ \hat{D}^{\dagger}_2(z,t_2,\Delta)\text{e}^{i(\Delta + \eta' z)(t-t_2)}.
 \end{align}
Using the region 2 boundary condition stated earlier, the optical field solution for region 2 is found in a similar fashion as for region 1:
\begin{align}
\hat{a}_2(z,t) &= \hat{b}_2(0,t)\text{e}^{\frac{\alpha z}{2}} + \hat{b}^\dagger_1(0,2t_2-t) \text{e}^{\frac{\alpha z}{2}} \notag \\
&\qquad \times \frac{\alpha \left[ 1 - \text{e}^{-(\alpha + i(\eta_1 - \eta' (t-t_2)))z}\right]}{\alpha +i (\eta_1 - \eta'(t-t_2))} 
\end{align}
where we have defined the following operators 
\begin{align}
\hat{b}_1(0,t) &\equiv \hat{a}_1(0,t) + i\alpha \int^z_0 dz' \text{e}^{\frac{\alpha z'}{2}} \hat{\mathcal{D}}_1(z',t) .\notag \\
\hat{b}_2(0,t) &\equiv \hat{a}_2(0,t) + i\alpha\int^z_0 dz' \text{e}^{-(\frac{\alpha}{2} + i(\eta_1 - \eta' (t +t_2)))z'} \notag \\
& \qquad\qquad\qquad\qquad \times\hat{\mathcal{D}}^\dagger_1(z',t-2t_2) \notag  
\end{align}
This completes the solution for region 2. It is noted that when the Stark fields are set to zero, the two pulse photon echo solutions are retained with efficiency $\eta_{2PE} = 4\sinh^2(\frac{\alpha z}{2})$ as expected.

We can now form the region 3 boundary condition, in a similar fashion to the region 2 boundary condition, namely $\hat{D}_3(z,t = t_4,\Delta) = \left[ \hat{D}^\dagger_2(z,t_4,\Delta) \right]^{\dagger}$.

The equations of motion that describe the dynamics of region 3 are
exactly those stated for region 1, Eqs.~\ref{equ:MBE1R1} and~\ref{equ:MBE2R1}. Hence the
optical solution for region 3 is identical in form to
Eq.~\ref{equ:a1solution}, with subscript $1\to 3$ and
$\hat{\mathcal{D}}_3(z,t) \equiv
\mathcal{F}^{-1}\left[\hat{D}_3(z,t_4,\omega)\text{e}^{-i\omega
    t_4}\right]$.  For balanced Stark fields, the output optical
solution in region 3 is
\begin{align} \label{equ:a3solution}
&\hat{a}_3(z,t) = \hat{b}_3(0,t) \text{e}^{-\frac{\alpha z}{2}} \;+\; \hat{b}^{\dagger}_2(0,2t_4-t) \text{e}^{-\frac{\alpha z}{2}} \notag \\
&\times \frac{\left[ 1-\text{e}^{(\alpha - i\eta' (t-t_4))z} \right]}{{\alpha} - i{\eta'}(t-t_4)} \;-\; \hat{b}_1(0,t-2t_4 + 2t_2)\text{e}^{-\frac{\alpha z}{2}}\notag \\
&\times \Bigg[ \alpha z + \alpha^2\Bigg( \frac{\text{e}^{2i\eta'(t - t_4)z} - 1}{2i\eta' (t-t_4)(\alpha - i\eta'(t-t_4))} \notag \\
& \qquad\qquad\qquad\qquad\qquad- \frac{\text{e}^{(\alpha + i\eta' (t-t_4))z} - 1}{|\alpha + i\eta' (t-t_4)|^2} \Bigg) \Bigg],
\end{align}
where  we have defined the operator
\begin{equation}
\hat{b}_3(0,t) = \hat{a}_3(0,t) + i\alpha \int^{z}_0 dz' \text{e}^{\frac{\alpha z'}{2}} \hat{\mathcal{D}}_1(z',t-2t_4+2t_2).\notag
\end{equation}

In the limit of large Stark shift, Eq.~\ref{equ:a3solution} becomes
\begin{align} \label{equ:a3solution2}
&\hat{a}_3(z,t) = \hat{a}_3(0,t) \text{e}^{-\frac{\alpha z}{2}} \;-\; \hat{a}_1(0,t-2t_4 + 2t_2)\alpha z \text{e}^{-\frac{\alpha z}{2}} \notag \\
&+ i\left[ \alpha(1 - \alpha z)\text{e}^{-\frac{\alpha z}{2}} \int^{z}_0 dz' \text{e}^{\frac{\alpha z'}{2}} \hat{\mathcal{D}}_1(z',t-2t_4 + 2t_2)\right].
\end{align}
The solution has three terms. The first term is the inevitable optical
vacuum input at the region 3 time boundary which decays as a function
of the optical depth. The second term is the HYPER echo which forms at
a time $t_5 = t_1 + 2t_4 - 2t_2$ with $\eta_{HPE} = (\alpha
z)^2\text{e}^{-\alpha z}$. This echo has a maximum efficiency of $54\%$
at $\alpha l = 2$. The third term contains the atomic degrees of
freedom. If counter-propagating $\pi$-pulses are used, causing the echo to be emitted in the backward direction, 100\% efficiency is possible (see appendix).

It can be seen that the Stark shifting field over region 2 eliminates
contributions from region 2 on the region 3 output. Taking the limit
of infinite Stark shift is physically equivalent to decoupling the
optical and atomic fields in the equations of motion in region 2, thus
`switching off' the collective atom-light interaction in this region
and reducing the noise on the output field. In the decoupled regime,
the atomic boundary condition at region 3 $(t = t_4)$ is found by
propagating the region 2 boundary condition $(t = t_2)$ forward in
time by $t_4 - t_2$. The result obtained for the output field in
region 3 is identical to Eq.~\ref{equ:a3solution2}.

\section{Echo demonstration}

\begin{figure}
  \centering
  \includegraphics[width=0.40\textwidth]{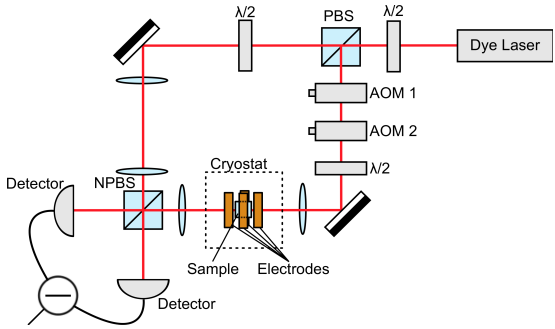}
\caption{\label{fig:expsetup}(Color online)
 Experimental setup. PBS - Polarizing beamsplitter, NPBS -
 non-polarizing beamsplitter, AOM - acousto optic modulator. A third
 AOM was used (not shown) before the interferometer to provide the
 required frequency shifts for repumping the praseodymium dopants to
 the required hyperfine states. 
 The  0.02\% \PRYSO crystal is mounted between four  electrodes in a
 quadrupolar arrangement that are used to apply the electric field
 gradients to the samples. 
 }
\end{figure}

The experimental setup used is shown in Fig.~\ref{fig:expsetup}, a
modified Coherent 699 dye laser with sub 5~kHz linewidth drives the
$^3$H$_4$ $-$ $^1$D$_2$ transition, at 605.977~nm, in a \PRYSO
crystal. The sample is oriented such that the laser
propagates along the crystal b-axis, and is linearly polarized along
the D$_2$-axis.

The crystal is surrounded by four electrodes in a quadrupole
arrangement that enabled an electric field gradient to be produced
along the direction of propagation. The electric fields used in region
1 varied from $\pm15$~V/cm over the optical path length, and were
directed parallel to the b-axis. Using the tabled value of the Stark
shift\cite{graf98}, this results in detunings of approximately
$\pm2$~MHz. The direction of the applied electric field was parallel
to the crystals b-axis.  The entire sample/electrode setup is mounted
inside a cryostat and cooled to $3.1\textrm{K}$.  Two acousto-optic
modulators (AOMs) were used to gate the laser beam to produce the
input pulses. The AOMs also introduce a net 10~MHz frequency shift,
enabling the light exiting the crystal to be detected with phase
sensitive heterodyne detection.

\begin{figure}
  \centering
  \includegraphics[width=0.4\textwidth]{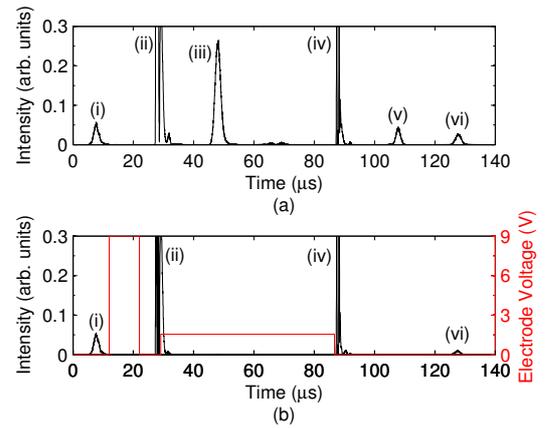}
  \caption{(Color online) An experimental hybrid photon echo trace (a) without and (b)
    with the electric field gradient applied. The two
    $\pi$-pulses saturated the detector.}
  \label{fig:echoseq}
\end{figure}

Figure \ref{fig:echoseq} shows the intensity of the light exiting the
sample during a hybrid photon echo sequence. Three optical pulses are
applied, the input pulse and two $\pi$-pulses. The input pulse is a
Gaussian with a full width at half maximum (FWHM) power of $1.8 \mu
s$, and the transmitted portion appears at (i). In the absence of the
Stark field, the first $\pi$-pulse (ii) causes a 2PE (iii). The 2PE is
rephased to form the hybrid photon echo at (vi). The two $\pi$-pulses
were not placed symmetrically about the 2PE so that the three pulse photon echo (3PE) at (v), due
to the input and two $\pi$-pulses, is separated in time from the HPE.

The amplitude of the pulses were calibrated by applying a
2PE sequence with the first pulse half the length of the
second. The lengths that maximised the size of the echo were taken as
being a $\pi/2$ pulse and $\pi$ pulse.

Applying the electric field to the sample causes the intensity of the
2PE to be reduced by $> 99\%$ through the mechanism explained
earlier. The 3PE also experiences a large reduction in intensity,
which can be explained using similar reasoning.  The HPE is still
formed albeit with reduced efficiency. We believe this reduction in
efficiency is most likely due to imperfect balancing of the broadening
applied before and after the first $\pi$-pulse.

\begin{figure}
  \centering
  \includegraphics[width=0.40\textwidth]{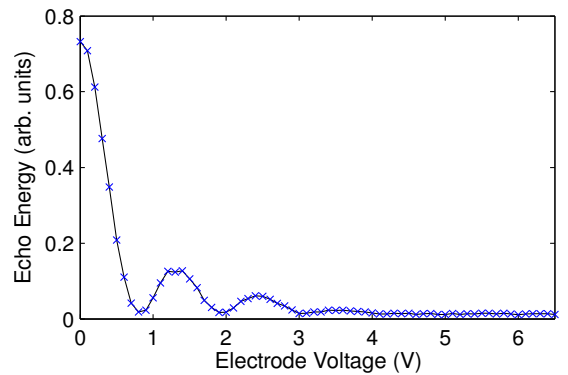}
  \caption{(Color online) \label{fig:echovfield} Disrupting the two
    pulse echo with
    the broadening field. The energy in the two pulse echo is plotted
    versus the magnitude of the electric field gradient applied to the
    sample.}
\end{figure}

Figure~\ref{fig:echovfield} shows the intensity of the 2PE
as the size of the broadening field is increased. The
electric field varies linearly along the propagation axis, 
resulting in a top-hat distribution of frequency shifts.  Thus the
phase shifts of the ions will have a top-hat distribution as the 2PE forms.
This leads to a sinc-squared behaviour in the energy of the 2PE, analogous to a single slit diffraction pattern.

Due to optical pumping distributing population amongst the hyperfine
levels \PRYSO exhibits spectral holeburning \cite{holi93}. Optical
repumping was used so that the number of ions and thus the optical
depth remained consistent between shots. The laser was first scanned over the spectral region of
interest to burn a wide spectral hole. This hole was then filled 
using laser light pulses detuned by a specific combination of the hyperfine
splittings. Varying the number of these repumping pulses allowed for
variation of the optical depth.

\begin{figure}
  \centering
  \includegraphics[width=0.40\textwidth]{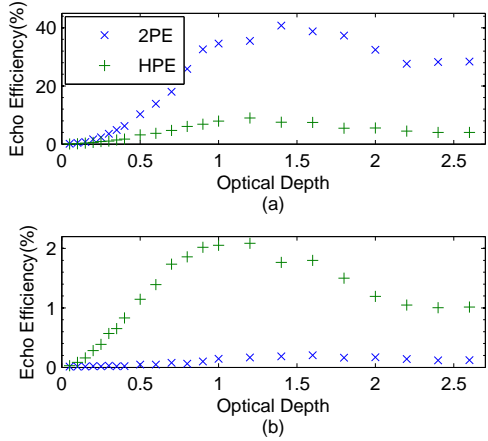}
\caption{\label{fig:odvefficiency} (Color online) Efficiency of the 2PE and HPE as the optical depth is varied. (a) Zero applied electric field, and (b) 9~V applied to electrodes. }
\end{figure}

The efficiency of the 2PE and HPE was measured over a range of optical
depths (see Fig.~\ref{fig:odvefficiency}).
The maximum efficiency for the 2PE occurs at an optical depth of around
1.5, while the maximum efficiency of the HPE is at an optical depth
of around 1.2. This is independent of whether the electric field was on or
off. The maximum 2PE efficiency reduces from 40\% to less than 0.3\%
when 9~V was applied to the electrodes.

\begin{figure}[t]
  \centering
  \includegraphics[width=0.40\textwidth]{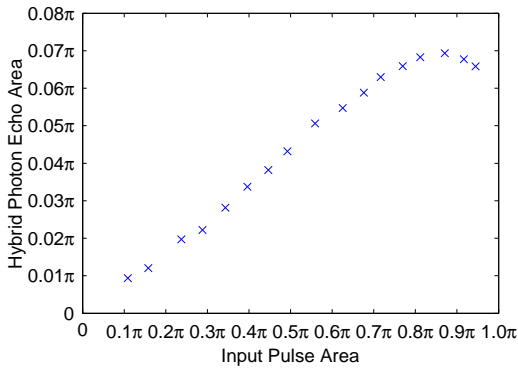}
\caption{\label{fig:inputoutput} (Color online) Area of the hybrid photon echo as the area of the input pulse is varied. 9V was applied to the electrodes.}
\end{figure}

In order to verify the linearity and determine the efficiency of the
HPE, the area of the input pulse was compared to the area of the
output pulse (see Fig.~\ref{fig:inputoutput}).  The length of the
input pulse was set to $1.8\mu s$ (FWHM), whilst the amplitude of the
pulse was altered by adjusting the radio frequency power applied to
AOM2.  The optical depth for this experiment was set to 1. As shown in
Fig.~\ref{fig:inputoutput}, the relationship between input pulse and
output pulse is linear for input pulses small compared to a $\pi$-pulse 
and then flattens off for more intense pulses.

\section{Noise Measurements}

To characterize the noise properties, the heterodyne signal of the
HYPER pulse sequence was recorded with no input pulse, both with and
without the electric field applied. The \PRYSO sample used for the
noise measurements was 4x4x20~mm (20~mm along the beam propagation
direction), doped at 0.005\%.  Before the initial $\pi$-pulse, 100~V is
applied to the two electrodes. In this alternate sample geometry
this results in an electric field varying from -30~V/cm to
+30~V/cm along the 20~mm axis, corresponding to a frequency shift
of approximately $\pm4$~MHz.  This voltage is decreased to 37.5~V for the longer
period between the $\pi$-pulses to maintain a zero net Stark shift
such that an input pulse would be rephased as a HPE.


The noise was calculated as follows. For each of $\sim$30~000 shots,
the time-domain signal was multiplied by the temporal envelope of the
would-be 2PE (FWHM of 1.8~$\mu$s, centered 15~$\mu$s after the first
$\pi$-pulse), numerically heterodyned (optical beat frequency here was
6~MHz) down to DC and summed. This quantifies the field amplitude in
the 2PE mode for each shot. Taking the variance across all shots
yields a value related to the average field intensity in this specific
mode. For normalization purposes, the shot-noise (noise level with local
oscillator on, but signal beam blocked) was measured as the variance in an
equivalent mode, but with the temporal envelope shifted to before the
first $\pi$-pulse.

\begin{figure}[h]
	\centering
	\includegraphics[width=1.0\columnwidth]{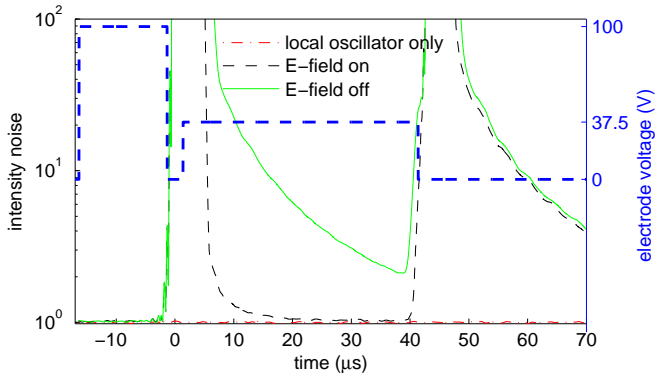}
        \caption{(Color online) Noise level at the heterodyne beat
          frequency between applied pulses and the local oscillator
          (6~MHz) without any electric field applied (solid green), and with
          the field applied (thin, dashed black).  Also shown is the shot noise
          level (dot-dashed red), and the electrode voltage profile (thick, dashed blue - with
          the scale on the right).  The $\pi$-pulses occur at 0 and
          42.7~$\mu$s, and the noise level in the period between the
          two pulses is seen to be dramatically reduced when the
          electric field is applied.  If the $\pi$-pulses were
          perfect, the noise following the latter pulse would also be
          close to the shot noise level.  For this data series,
          ${\alpha}L=0.15$.  }
\label{fig:NoiseFig}
\end{figure}

At an optical depth of 0.15, the quadrature variances in the echo mode
\emph{without} the electric field applied were measured as 8.64$\pm$8\%
normalized to the shot-noise level.  With the electric field applied,
the noise reduces approximately two orders of magnitude to 1.045$\pm$4\%,  a level indistinguishable from the vacuum.
Figure~\ref{fig:NoiseFig} shows the averaged time dependence of the
noise in the spectral mode of the echo. The experiment was repeated
for larger optical depths of 0.36 and 0.98, where the noise was
similarly reduced by factors of 150 and 160 respectively.

There are three mechanisms through which the applied field gradient
decreases the noise.  Firstly, the gain feature created by the initial
$\pi$-pulse is broadened in frequency, and therefore the number of
excited ions generating amplified spontaneous emission in the original
frequency window is reduced. Secondly, there are ground-state ions
Stark-shifted into the detection frequency window which increases the
absorption. Assuming a perfect $\pi$-pulse with a 1~MHz width, and an
optical depth of 1, the expected reduction of noise due to these two
effects is only $\sim$5. However, in practice the $\pi$-pulse is far
from perfect and most of the ions in the ensemble see a pulse area
greater or less than $\pi$ (due largely to the Gaussian spatial
profile of the pulse intensity). This and structure on the
inhomogeneous line results in a large free induction decay (FID). The
third and most significant effect of the applied electric field is the
reduction of this FID.

\begin{figure}
  \centering
  \includegraphics[width=0.5\textwidth]{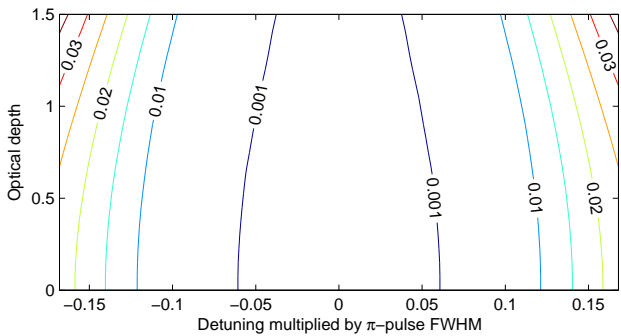}
  \caption{\label{fig:semiclass}(Color online) The fraction of  atoms
    in the excited state shortly after two $\pi$-pulses. The medium is
    initially in the ground state then two Gaussian shaped
    $\pi$-pulses are sequentially applied to the medium. The pulses
    were separated by 20 times the FWHM of their amplitudes. The plot
    is generated from the solution of the one dimensional semi-classical Maxwell-Bloch equations.}
\end{figure}

\section{Discussion and Classical Simulation}

There are two areas where these initial experiments need to be
improved before the proposed memory can fulfil its theoretical
promises. The demonstrated memory is of low efficiency and noisy. The
noise measurements, Fig.~\ref{fig:NoiseFig}, show that the broadening field is
successful in decoupling excited atoms from the output field while they
are inverted, but there is excess noise in the time region where the light would be
recalled from the memory. 

The predominant reason for this noise is the imperfect $\pi$-pulses
resulting in random FIDs due to structure in the
inhomogeneous line. In systems that have long lived spectral holes
like \PRYSO this structure is hard to avoid. A four-level echo
protocol provides a way around this problem \cite{beav11}. Experiments
involving Tm$^{3+}$:YAG, which has short lived holes, have achieved $\pi$-pulses 
without measurable random FIDs
\cite{ledingham_in_prep}.

Even in the absence of structure on the inhomogeneous line, this HYPER quantum
memory scheme is still reliant on doing $\pi$-pulses well, but less
sensitively. In the output region the solution for the time domain is
an expression like Eq.~\ref{equ:a1solution}, but with the subscript 1
replaced by 3, for region 3. Here $\hat{a}_3(0,t)$ represents the
light incident on the crystal during the recall of the light (usually
the vacuum). The term that includes $\mathcal{\hat{D}}_3(z,t)$
represents that light emitted from the sample due to any excitation in
the crystal after the second $\pi$-pulse. In the case of perfect
$\pi$-pulses and vacuum input, $\hat{\mathcal{D}}_3(z,t)$ will
represent ground state atoms. In the case of imperfect $\pi$-pulses
the atoms will have some unwanted excitation resulting in noise. The
ability to apply accurate $\pi$-pulses has already received
careful study \cite{zafa07,rugg10}. Semi-classical 1-D numerical
calculations, presented in Fig.~\ref{fig:semiclass}, show that in the
case of no transverse variation in the optical intensity, atomic population very
close to the ground can be achieved over a significant fraction of the
bandwidth of the pulses. This in turn means that the amount of noise
in the output would be reduced to much lower than one photon per
spatio-temporal mode. The fact that there will be spontaneous emission while the atoms are excited will be another noise process, but like the effect of imperfect $\pi$-pulses this noise would not be phase matched with the echo. As long as the population is predominantly in the ground state when the echo is retrieved, the amount of noise will be much less than one photon per spatio-temporal mode.

\section{Conclusion}

We propose and present initial experiments for a new quantum memory
protocol. The proposal uses strong rephasing pulses rather than
structure in the inhomogeneous line made with spectral
holeburning. Noise associated with echoes forming in the excited state
is avoided by using electric field gradients to generate artificial
inhomogeneous broadening while the media is in the excited state. We
present a demonstration of these echoes as well as initial noise
measurements. In these experiments the echoes were noisy due to random
FIDs arising from structure on the inhomogeneous line rather than
amplified spontaneous emission from the excited state atoms. In order to make
the memory quiet these random FIDs must be avoided. Furthermore to ensure the atoms form the echo while close to the ground state accurate
$\pi$-pulses will be required. Both of these steps should be
technically feasible.

\textit{Note added:} Recently, a new proposal has been developed that uses some of these ideas \cite{rose}. This new proposal is attractive in that it removes the need for auxiliary broadening fields.

\section{Acknowledgments}

We thank the referee who suggested we look at backward retrieval which led to the material in the appendix.

DLM, PML, WRN and JJL were supported by the New Zealand Foundation for Research Science and Technology under Contract No.\ NERF-UOOX0703.

SEB, MPH, and MJS were supported by the Australian Research Council Centre of Excellence for Quantum Computation and Communication Technology (Project No. CE11E0096).

\appendix*

\section{Backward Retrieval}

Here we show that similar to other proposals \cite{mois01,afze09,rose}, in the limit of large optical depth the memory has perfect efficiency when the echo is retrieved from the sample in the backward direction.

Here we consider a sample that lies from $z=0$ to $z=L$. Equations~\ref{equ:MBE1R1} and~\ref{equ:MBE2R1} are equations of motion for the atomic ($D_1$) and optical ($a_1$) modes propagating in the forward direction, toward larger values of $z$. For clarity the labels $D_f$  and $a_f$ will be used for these quantities. The equations of motion for the counter-propagating, `backward' fields are the same as the forward case except for a minus sign in Eq.~\ref{equ:MBE2R1}
\begin{equation}
    \frac{\partial}{\partial z}  \hat{a}_{b}(z,t) = {\bf -}\frac{i \alpha}{2 \pi} \int_{-\infty}^{\infty} d\Delta \;   \hat{D}_{b}
\end{equation}

The treatment in this appendix is semi-classical. This simplifies the expressions greatly because the amplitudes of modes with no excitation can be set to zero, and then ignored. A quantum treatment will yield the same results, but with the appropriate addition of vacuum modes to preserve the commutation relations in the presence of loss.

We consider an input field $a_f(0,t) = a_\text{in}(t)$ that is zero for $t>0$, and atomic modes that are initially in the ground state.
In this case, we can see from Eqs.~\ref{equ:D1solution} and~\ref{equ:a1solution} that the forward propagating atomic modes at $t=0$ are given by
\begin{eqnarray}
  \label{eq:jevon1}
  D_f(z,t=0,\Delta) &=& i \int_{-\infty}^{\cancelto{\infty}{0}}a_f(0,t')\text{e}^{-\frac{\alpha z}{2}}\text{e}^{-i\Delta t'}\,dt'\nonumber\\
  &=&i\text{e}^{-\frac{\alpha z}{2}} a_\text{in}(\omega=\Delta).
\end{eqnarray}
Here $a_\text{in}(\omega)$ is the Fourier transform of $a_\text{in}(t)$. The upper limit on the integral can be changed to infinity because the input field is zero for $t>0$.

We reduce the dynamics during the rephasing period, the time period between when the broadening field is first turned on and the second rephasing pulse, to an instantaneous operation at $t=0$. For the case of counter-propagating $\pi$-pulses this has the effect
\begin{equation}
  D_b(z,t=0,\Delta) = D_f(z,t=0,\Delta)\text{e}^{-i\Delta t_d}
\end{equation}
The time delay, $t_d$, is the difference between how long the atoms are in the excited state and the ground state, during the rephasing period. Analogous to Eq.~\ref{equ:a1solution} we have for $t>0$

\begin{equation}
{a}_b(z,\omega) = {a}_b(L,\omega)\text{e}^{-\frac{\alpha (L-z)}{2}} -
i\alpha \int^L_z dz' \text{e}^{-\frac{\alpha}{2}(z'-z)}D_b(z',0,\omega) 
\end{equation}

The first term on the right-hand side can be ignored because no light is incident on the $z=L$ end of the crystal during the retrieval process. The integral can easily be computed, resulting in a value for $a_\text{out}(\omega) = a_b(0,\omega)$ of 
\begin{equation}
 -a_\text{in}(\omega)\left(1-\text{e}^{-\alpha L}\right)e^{-i\omega t_d}
\end{equation}
or alternatively
\begin{equation}
a_\text{out}(t) =  -a_\text{in}(t-t_d)\left(1-\text{e}^{-\alpha L}\right),
\end{equation}

leading to an efficiency that varies with optical depth as $(1-\text{e}^{-\alpha L})^2$. This is equal to 1 for large $\alpha L$.


\end{document}